\documentclass{PoS}
\PoS{PoS(LAT2005)325}
\FullConference{XXIIIrd International Symposium on Lattice Field Theory\\
         25-30 July 2005\\
         Trinity College, Dublin, Ireland}

\usepackage{graphicx}
\newcommand{\beq}{\begin{displaymath}}
\newcommand{\eeq}{\end{displaymath}}
\newcommand{\bea}{\begin{eqnarray}}
\newcommand{\eea}{\non\end{eqnarray}}
\renewcommand{\d}{\delta}
\renewcommand{\l}{\lambda}

\newcommand{\m}{\mu}
\renewcommand{\r}{\rho}

\newcommand{\s}{\sigma}

\newcommand{\oh}{\frac{1}{2}}

\newcommand{\dg}{\dagger}
\newcommand{\non}{\nonumber}

\title{Localized eigenmodes of the covariant lattice Laplacian%
\addtocounter{footnote}{1}\thanks{Our research is supported in part
by the U.S. Department of Energy under Grant No.\ DE-FG03-92ER40711
(J.G.), the Slovak Science and Technology Assistance Agency under
Contract No.\ APVT-51-005704 (\v{S}.O.), grants RFBR 04-02-16079,
RFBR 05-02-16306-a (M.I.P.),  grants RFBR 05-02-16306-a, RFBR
05-02-17642, and an Euler-Stipendium (S.N.S.).}}
\ShortTitle{Localized eigenmodes of the covariant lattice Laplacian}
\author{\addtocounter{footnote}{-2}\speaker{J.\ Greensite}\\
        The Niels Bohr Institute, Blegdamsvej 17, DK-2100 Copenhagen \O, Denmark;\\
        Physics and Astronomy Dept., San Francisco State University,
        San Francisco, CA~94117, USA\\
        E-mail: \email{greensit@stars.sfsu.edu}}
\author{\v S.\ Olejn\'\i k\\
        Institute of Physics, Slovak Academy of Sciences,
        SK--845 11 Bratislava, Slovakia\\
        E-mail: \email{stefan.olejnik@savba.sk}}
\author{M.\ I.\ Polikarpov, S.\ N.\ Syritsyn\\
        ITEP, B.\ Cheremushkinskaya 25, Moscow 117259, Russia\\
        E-mail: \email{polykarp@itep.ru}, \email{syritsyn@itep.ru}}
\author{V.\ I.\ Zakharov\\
        Max-Planck Instit\"ut f\"ur Physik, F\"ohringer Ring 6,
        D-80805 Munich, Germany\\
        E-mail: \email{xxz@mppmu.mpg.de}}
\abstract{We study numerically the eigenmode spectrum of the
covariant lattice Laplacian, in the fundamental SU(2) color group
representation. It is found that eigenmodes at the lower and upper
ends of the spectrum are localized, and that the localization volume
scales. In contrast, the eigenmodes of the lattice Faddeev--Popov
operator are all extended rather than localized (as required for
confinement) despite the similarity of the kinetic and
Faddeev--Popov operators.}

\begin{document}

\section{Introduction}

    It is well-known from condensed-matter physics that electron
propagation in a periodic potential is described by Bloch waves,
which are extended, plane-wave-like states. However, when disorder
is introduced into the potential, low-lying electron eigenstates
become exponentially localized, as shown long ago by Anderson. This
is an interference effect due to multiple scattering, rather than
ordinary bound state formation in a single potential well. When the
energy of the highest localized state (the ``mobility edge'')
exceeds the Fermi energy, the material is an insulator.

    Recently, localization in the lowest eigenmodes of the lattice Dirac
operator has been intensely studied, in the hope that it might shed
light on properties and dimensionality of important underlying
structures in the QCD vacuum. It was found that:
\begin{list}{$\bullet$}{\topsep3pt\parsep0pt\itemsep3pt}
\item
   Wilson--Dirac fermions have a low-lying spectrum of localized
eigenmodes in certain regions of the phase diagram~\cite{Maarten};
\item
   low-lying modes of the Asqtad fermion operator, although extended,
seem to concentrate on lattice sub-volumes of dimensionality < 4
\cite{MILC};
\item
    low-lying modes of the overlap Dirac operator are localized on volumes
which shrink as a power of the lattice spacing $a$~\cite{overlap}.
\end{list}

    Some questions naturally arise: If fermionic operators are picking
up signals of lower-di\-men\-sion\-al substructure, is there any
relation to, e.g., center vortex sheets or monopole worldlines? Can
one find indications of lattice-scale 2-brane structures, along the
lines suggested by Zakharov \cite{Valya}?

    In the present study we concentrate on the following questions:
\begin{list}{$\bullet$}{\topsep3pt\parsep0pt\itemsep3pt}
\item
   Is localization/concentration unique to Dirac operator eigenmodes,
or is it found in other lattice kinetic operators, e.g. the
Faddeev--Popov and covariant Laplacian operators?%
\footnote{Note that the covariant Laplacian is not the square of the
Dirac operator.  (This is only true for the free theory.) That means
that the eigenmodes of the covariant Laplacian need not be directly
related to the eigenmodes of the Dirac operator; they might have
completely different localization properties.}
\item
   If so, is there any connection to confinement?
\end{list}
Here we sketch a subset of our results, other can be found in
\cite{GOPSZ,Sergey}.

\section{Signals of localization}

    The covariant lattice Laplacian in the $j$-th representation of the
SU(2) gauge group is
\begin{equation}
      \triangle_{xy}^{ab} = \sum_{\m} \left[ U^{ab}_\m(x) \d_{y,x+\hat{\m}}
         + U^{\dg ab}_\m(x-\hat{\m}) \d_{y,x-\hat{\m}}  - 2 \d^{ab} \d_{xy} \right]
\end{equation}
(color indices $a, b$ run from 1 to $2j+1$) and we are interested in
the low-lying eigenmodes $\phi_n^a(x)$ satisfying the eigenvalue
equation
$-\triangle_{xy}^{ab} \phi_n^b(y) = \l_n \phi_n^a(x).$

    As probes of localization, we use two quantities:
\begin{list}{$\bullet$}{\topsep3pt\parsep0pt\itemsep3pt}
\item
    The \textbf{\emph{Inverse Participation Ratio}} ($IPR$) of the $n$-th
eigenmode, defined by
\begin{equation}
          IPR_n = V \left\langle \sum_x \r_n^2(x) \right\rangle,\qquad
\mbox{where} \qquad
         \r_n(x) = \sum_a \left\vert \phi^a_n(x) \right\vert^2
\end{equation}
is the normalized eigenmode density. If the eigenmode is extended
over the whole lattice, $\rho\approx 1 / V$, and $IPR={\cal{O}}(1)$.
In contrast, for an eigenmode localized in a volume~$b$ (i.e.\
$\rho\approx 0$  except in a region of volume $b$), $IPR\approx
1/b$.
\item
    The \textbf{\emph{Remaining Norm}} ($RN$), defined in the
following way: Sort $\rho(x)$ into a one-dimensional array $r(k),
k=1,2,\dots,V$, with $r(k) \ge r(k+1)$. Then the $RN$ is
\begin{equation} RN(K) = 1 - \sum_{k=1}^K r(k). \end{equation}
The $RN$ is the amount of total norm ($=1$) remaining after counting
contributions from the $K<V$ subset of sites with largest $\rho(x)$.
\end{list}
%

\section{Results for the fundamental ($j=1/2$) representation}

\begin{figure}[b!]
\begin{center}
\begin{tabular}{c p{0.01\hsize} c}
  \includegraphics[width=0.45\hsize]{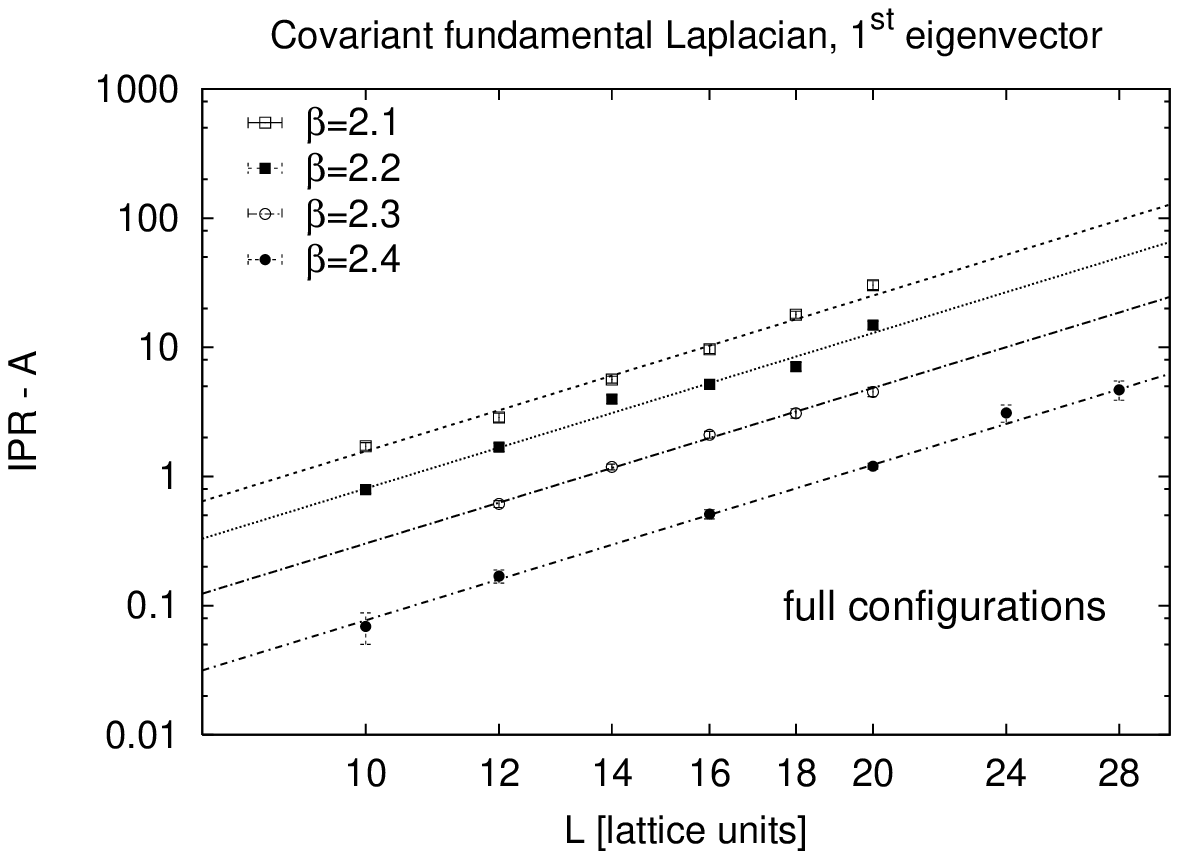}& &
  \includegraphics[width=0.45\hsize]{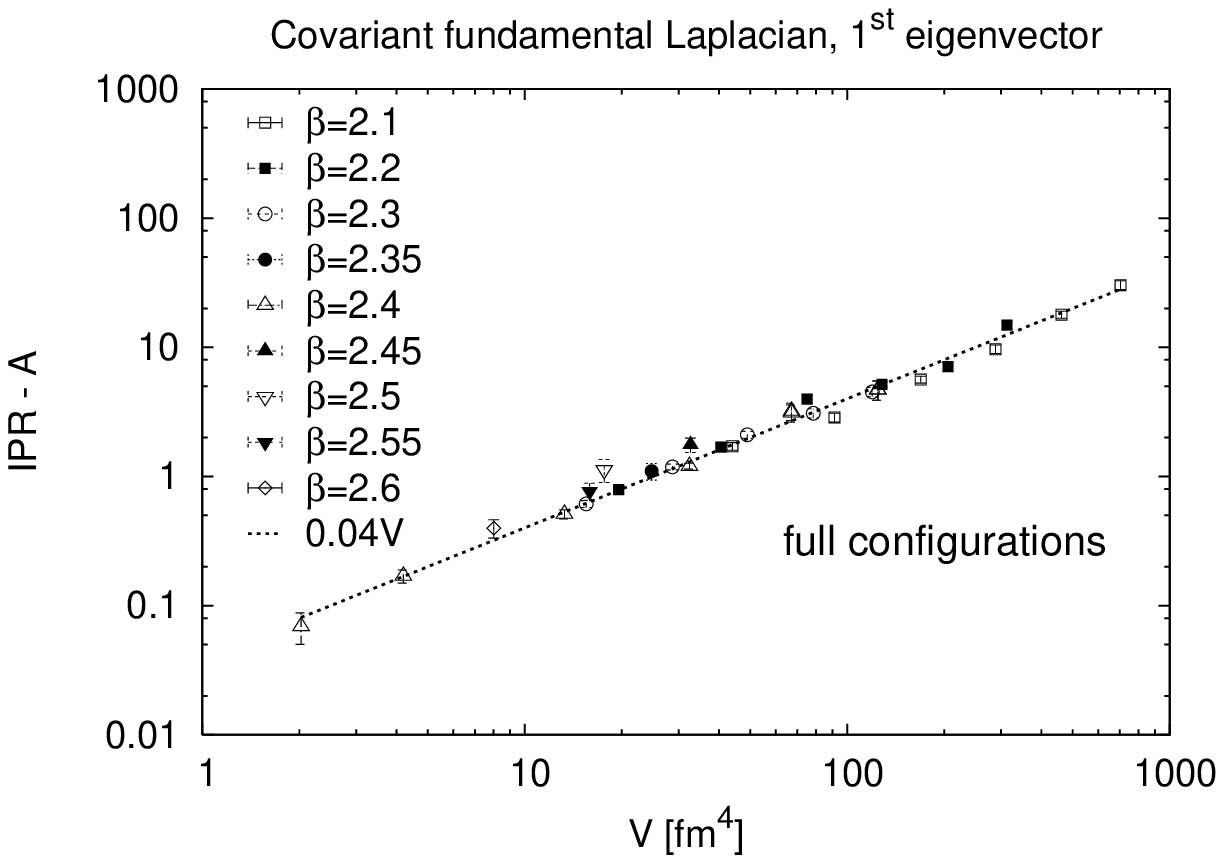}\\
\end{tabular}
\end{center}
\caption{[Left] Log-log plot of $IPR-A$ vs lattice length $L$, at
various $\beta$. The lines show a fit to $IPR = A + L^4/b$. [Right]
Log-log plot of $IPR-A$ vs physical volume $V=(La)^4$.}
\label{iprf2}
\end{figure}

    \textbf{\emph{Evidence of localization.}} We calculated the average
$IPR$ of the lowest-lying eigenmode at various $\beta$ values, and
fit the data at each coupling to $IPR=A+L^4/b$. Fig.~\ref{iprf2}
(left part) shows the data in a log-log plot. It is quite clear that
$(IPR-A)$ is proportional to the lattice volume, indicating that the
eigenmode is localized in a 4-volume. In Fig.~\ref{iprf2} (right) we
plot $(IPR-A)$ vs the lattice volume in physical units,
$V~{\rm[fm^4]}=Va^4(\beta)=(La^4(\beta))$, where $a(\beta)$ denotes
the lattice spacing (in fm) at coupling $\beta$. The data fall
roughly on a straight line, which implies that the localization
4-volume $(ba^4)$ is constant in physical units,
about~$(2.3~\mbox{fm})^4$.

    \textbf{\emph{Locating/removing center vortices.}} The next question
is whether the localization is due to some confining disorder of the
lattice configuration. Therefore we computed the spectrum of the
covariant Laplacian in the center-projected and vortex-removed
configurations. We first fixed lattice configurations into the
(direct) maximal center gauge (MCG) and obtained ``vortex-only''
configurations by center projection~\cite{DMCG}. The
``vortex-removed'' configurations were then obtained by multiplying
the lattice configuration in MCG by the center-projected
one~\cite{dFE}. Vortex removal is quite a minimal change -- only the
action at P-vortex plaquettes (plaquettes equal to $-1$ after center
projection) is changed, and the density of those plaquettes drops
exponentially with $\beta$. How is localization affected?

\begin{figure}[t!]
\begin{center}
\begin{tabular}{c p{0.01\hsize} c}
  \includegraphics[width=0.45\hsize]{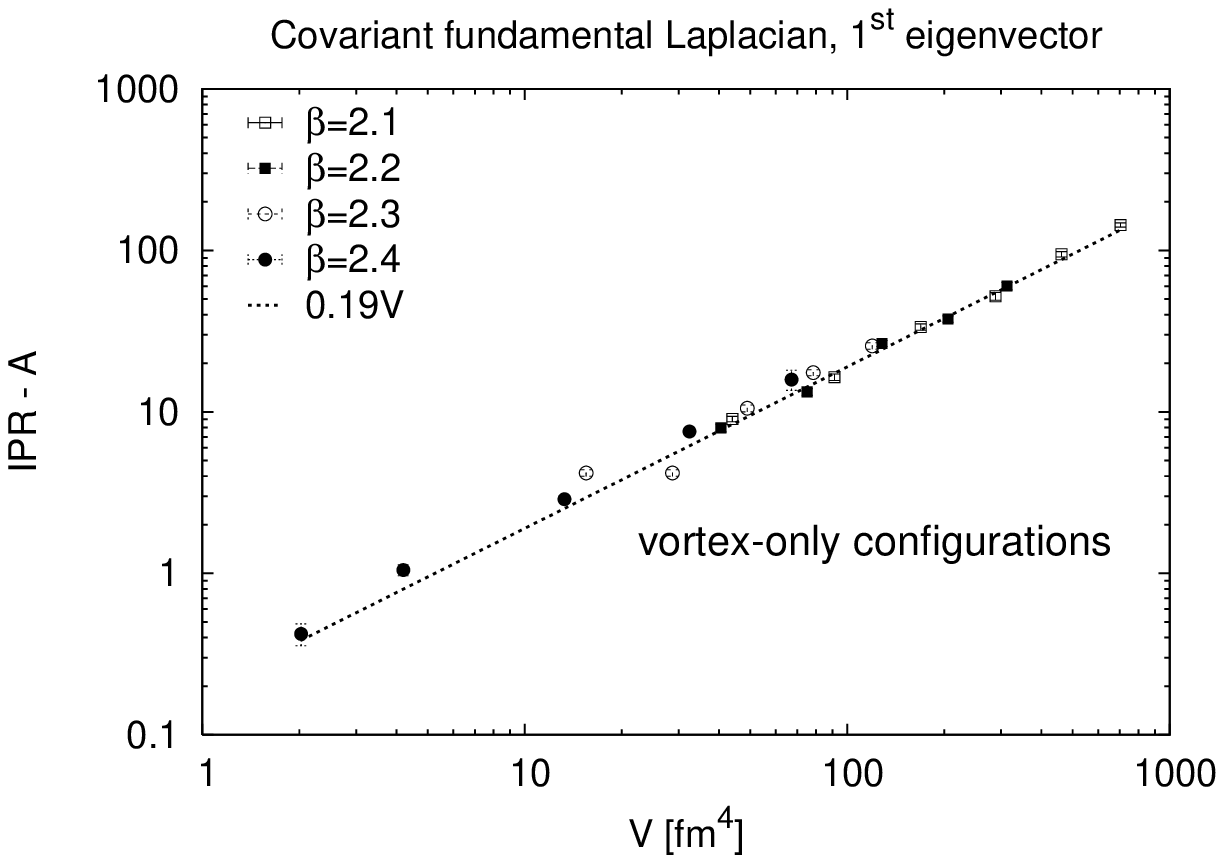}& &
  \includegraphics[width=0.45\hsize]{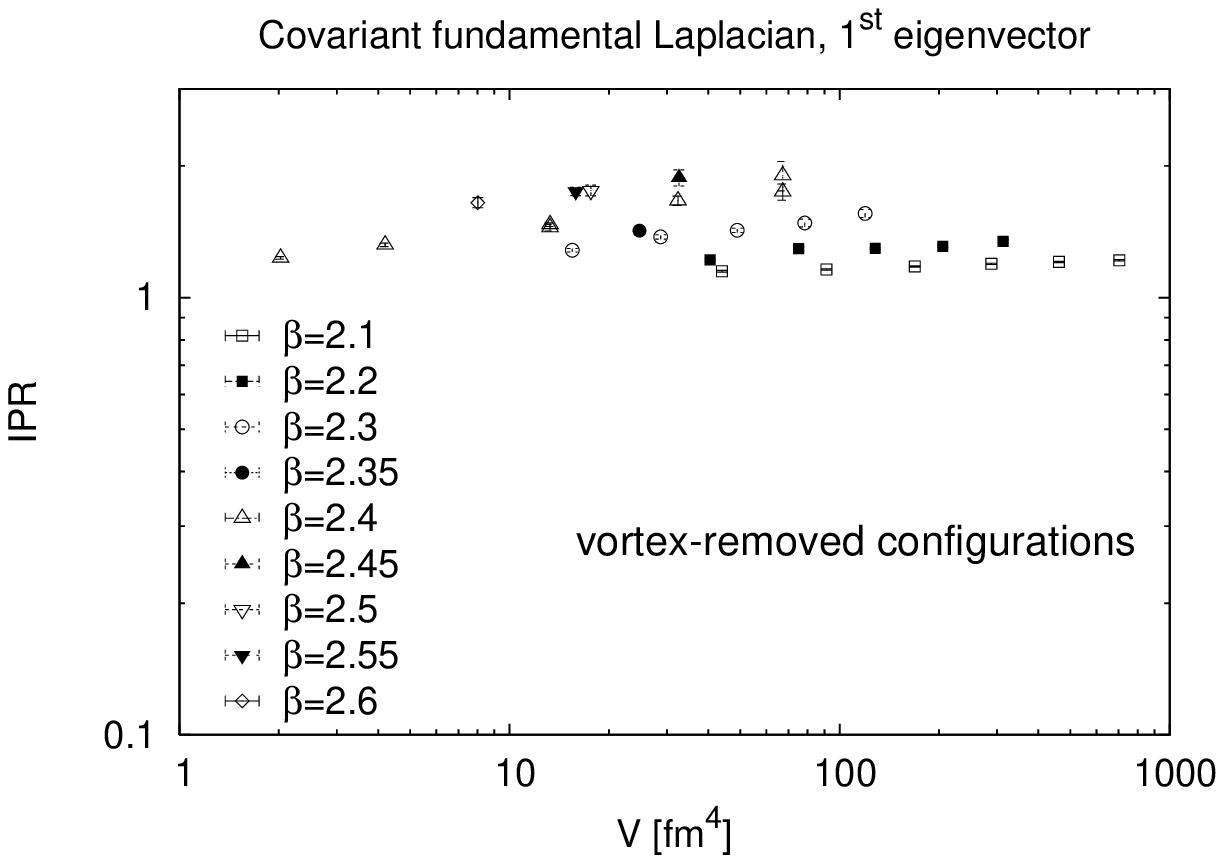}\\
\end{tabular}
\end{center}
\caption{Log-log plot of $IPR-A$ vs physical volume in vortex-only
configurations [left] and vortex-removed configurations [right].}
\label{iprp_ipr0}
\end{figure}

    Results are shown in Fig.~\ref{iprp_ipr0}. We find a somewhat
greater localization (left part) in vortex-only configurations.
$b_{phys}$ is reduced in this case to about $(1.5~\mbox{fm})^4$. The
situation drastically changes in vortex-removed configurations, the
eigenmodes turn out to be extended (see Fig.~\ref{iprp_ipr0},
right).

    The same conclusion follows from examining the Remaining Norm (see
\cite{GOPSZ}): For the unmodified configurations, the RN curve
becomes slightly broader with increasing lattice volume, but the
$RN$ data seem to converge to a limiting curve at the largest
volumes. The convergence is clearer in vortex-only configurations,
where almost all data fall on essentially the same curve. For the
vortex-removed data, the curve broadens as the volume increases;
again, this means no localization without center vortices.

\begin{figure}[b!]
\begin{center}
\begin{tabular}{c p{0.01\hsize} c}
  \includegraphics[width=0.45\hsize]{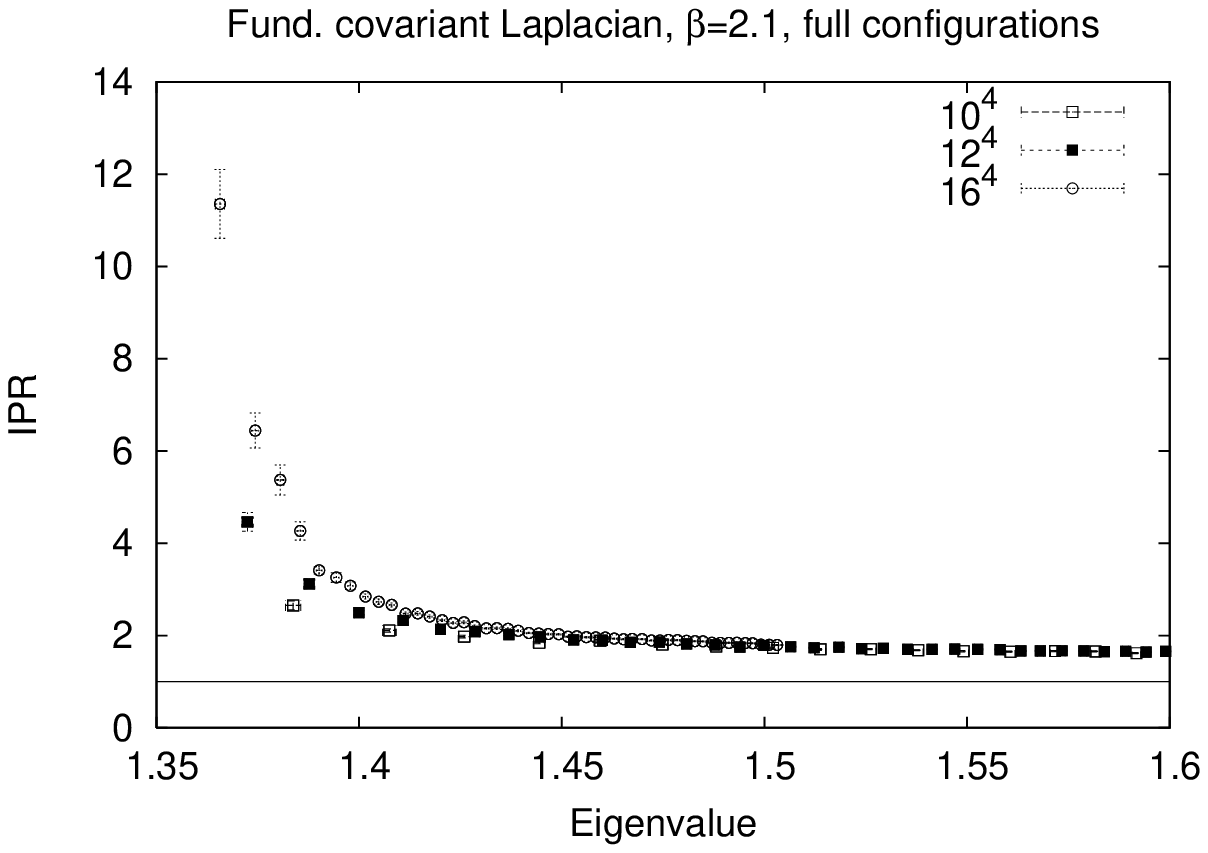}& &
  \includegraphics[width=0.45\hsize]{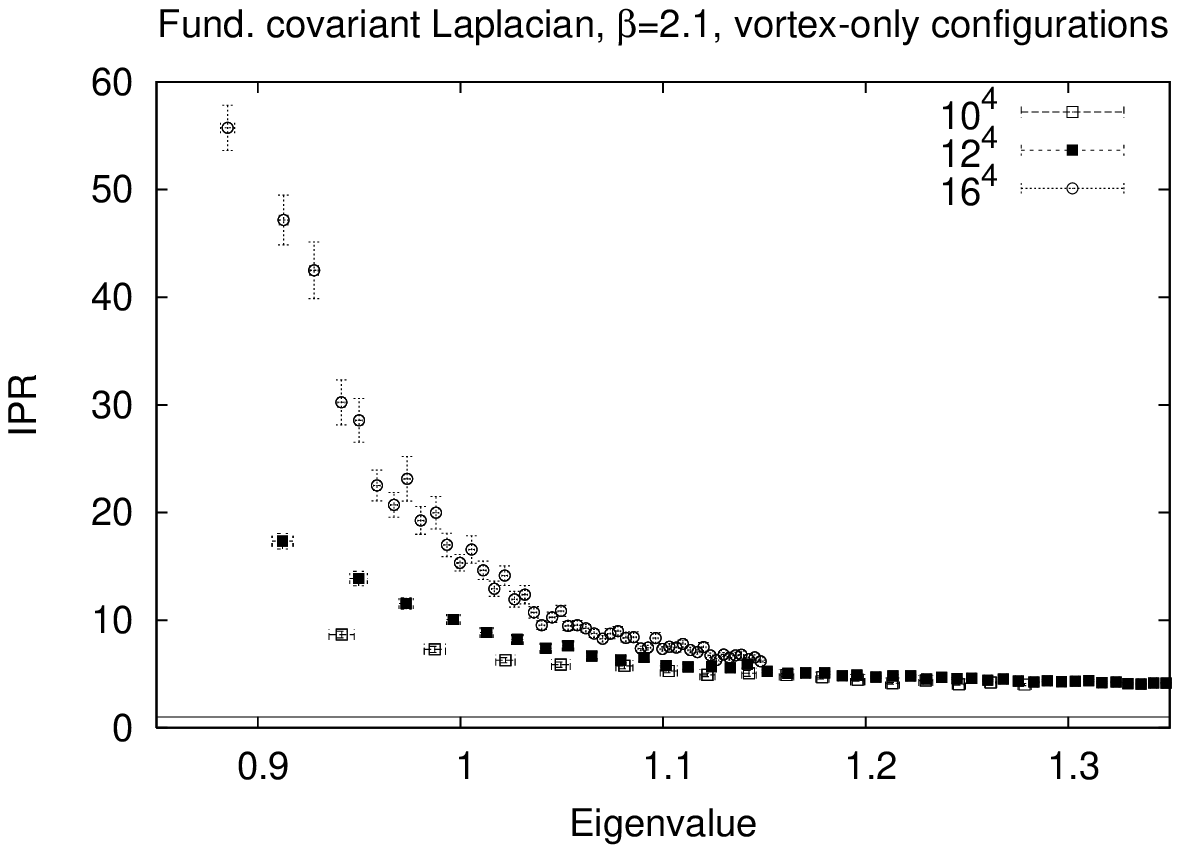}\\
\end{tabular}
\end{center}
\caption{IPRs of the first 100 eigenmodes in full [left] and
vortex-only configurations [right].} \label{ipr_vs_lambda_2p1}
\end{figure}

    \textbf{\emph{Mobility edges.}} Not all eigenmodes of the covariant
Laplacian in the fundamental representation are localized, only the
low- and high-lying modes.%
\footnote{If $\phi(x)$ is an eigenmode, so is $\phi'(x) =
(-1)^{\sum_\m x_\m} \phi(x).$ One can then show that $IPR_n =
IPR_{n_{max}- n + 1}$. Localization at the lower end of the spectrum
implies localization at the upper end~\cite{GOPSZ}.} The bulk of
states are extended. This is shown in Fig.~\ref{ipr_vs_lambda_2p1}
for full and vortex-only configurations. One could estimate (``by
the eye'') the mobility edge at $\beta=2.1$ to be around
$\lambda=1.45$ in the former, and about $1.15$ in the latter case.

\section{The Faddeev--Popov Operator}

    We have argued that localization of the low-lying states of scalar
particles implies a mass gap, i.e.\ the bare scalar mass cannot be
adjusted to zero and scalar field correlators cannot be long-range
(see \cite{GOPSZ}). A check of the logic of the argument is provided
by the Faddeev--Popov (FP) operator in Coulomb gauge. On the one
hand, it looks similar to the covariant Laplacian (in $D=3$
dimensions):
${\cal{M}} = - \vec{\nabla} \cdot \vec{\cal{D}}(A)$
[${\cal{D}}(A)$ is the covariant derivative]. One might therefore
expect that the low-lying eigenvalues are localized. On the other
hand, the Coulomb energy of a given charge distribution is
$$
       H_{coul} = \oh \int d^3x d^3y\; \rho^a(x)
               K^{ab}(x,y;A) \r^b(y),\qquad
\mbox{where}\qquad
       K^{ab}(x,y;A) = \left[ {\cal{M}}^{-1}
       (-\nabla^2) {\cal{M}}^{-1} \right]^{ab}_{xy},
$$
so if ${\cal{M}}^{-1}$ is short-range, the Coulomb potential $\sim
K(x,y)$ is short range as well.

\begin{figure}
\begin{center}
\includegraphics[width=0.45\hsize]{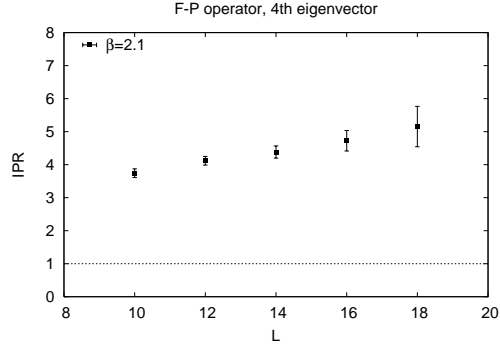}
\caption{IPR of the lowest nontrivial eigenmode of the FP operator.}
\label{FP_iprf}
\end{center}
\end{figure}

    But in fact, we know that asymptotically \cite{GO}
$ \langle K(x,y,A) \rangle \sim \s_{coul} |x-y|, $
where $\sigma_{coul}\approx 3\sigma$, so it must be that the
low-lying eigenmodes of the FP operator are not localized. The
$IPR$'s for the lowest nontrivial eigenmode of the FP operator at
$\beta=2.1$ are shown in Fig.~\ref{FP_iprf}. Despite the similarity
to the covariant Laplacian, we see no apparent localization -- in
agreement with Gribov--Zwanziger Coulomb-gauge confinement
scenario~\cite{horizon} and the claim formulated at the beginning of
this section.

\section{What feature of Laplacian eigenmodes is crucial to confinement?}

    If localization is not crucial to confinement, then what feature is? Let
us  suppose we have a dynamical scalar field $\phi(x)$ in the
Lagrangian. Consider the Euclidean propagator in Coulomb gauge
$ D(t) = \langle \phi^{\dg a}(t) \phi(0) \rangle
     =\sum_k c_k e^{-E_k t}, $
where the $E_k$ are the excitation energies of (non-singlet) states
with the quantum numbers of the scalar.  These should be infinite,
in a confining theory. Then, in the quenched approximation,
\begin{equation}
      D(t) = \sum_n \left\langle {\phi^{\dg a}_n(t)
           \phi^a_n(0) \over   \lambda_n + m_0^2} \right\rangle
       = 0,\qquad
       \mbox{where $m_0^2$ is the bare mass.}
\end{equation}
The natural implication is that
$ G_n(t) \equiv \left\langle \phi^{\dg a}_n(t)
           \phi^a_n(0)\right\rangle = 0 $
for any eigenmode, localized or extended, and any time difference
$t>0$.

    This we indeed find numerically in the confining phase, i.e.
$G_n(0,t) \approx 0$, $G_n(\vec{x},0) \ne 0$.
When center vortices are removed, or in the Higgs phase, then also
$G_n(0,t)\ne 0$.  This is a consequence of spontaneous breaking of
the remnant global gauge symmetry that exists in Coulomb
gauge~\cite{GOZ}.

\section{Conclusions}

    In the fundamental ($j=1/2$) representation of SU(2) we find that
the low lying eigenmodes of the covariant Laplacian operator are
localized, and that
\begin{list}{$\bullet$}{\topsep3pt\parsep0pt\itemsep3pt}
  \item $ba^4$  is $\beta$-independent:
  localization volume is fixed in physical units;
  \item the same is true for ``vortex-only''
  configurations; localization volume is smaller;
  \item localization disappears when vortices are removed;
  \item there are no localized eigenmodes for the Faddeev--Popov operator,
  as required in Coulomb-gauge confinement scenarios ~\cite{horizon}.
\end{list}
One might suppose that the situation will be much the same for
higher group representations.  In fact there are some surprises,
which were discussed in Sergey Syritsyn's talk~\cite{Sergey}.


\end{document}